\documentclass[aip,prl,reprint,twocolumn,superscriptaddress]{revtex4}
\usepackage{graphicx}
\usepackage{dcolumn}
\bibstyle{apsrev}	
\usepackage{color} 
\usepackage[normalem]{ulem}

\begin{document}

\title{Strong-coupling  charge density wave  in a one-dimensional topological metal}

\author{Philip Hofmann}
\affiliation{Department of Physics and Astronomy, Interdisciplinary Nanoscience Center (iNANO), Aarhus University, 8000 Aarhus C, Denmark}
\author{Miguel Ugeda}
\affiliation{CIC nanoGUNE,  20018 San Sebasti\'an-Donostia, Spain}
\affiliation{Ikerbasque, Basque Foundation for Science, 48011 Bilbao, Spain}
\author{Antonio J. Mart\'{\i}nez-Galera}
\affiliation{Dept. F\'{\i}sica de la Materia Condensada, Universidad Aut\'onoma de Madrid, Madrid 28049, Spain}
\author{Anna Str\'{o}\.{z}ecka}
\affiliation{Institut f\"{u}r Experimentalphysik, Freie Universit\"{a}t Berlin, 14195 Berlin, Germany}
\author{Jose M. G\'omez-Rodr\'{\i}guez}.
\affiliation{Dept. F\'{\i}sica de la Materia Condensada, Universidad Aut\'onoma de Madrid, Madrid 28049, Spain}
\affiliation{Instituto Nicolas Cabrera, Universidad Aut\'onoma de Madrid, 28049 Madrid, Spain}
\affiliation{Condensed Matter Physics Center (IFIMAC), Universidad Aut\'onoma de Madrid, 28049 Madrid, Spain}
\author{Emile Rienks}
\author{Maria Fuglsang Jensen}
\affiliation{Department of Physics and Astronomy, Interdisciplinary Nanoscience Center (iNANO), Aarhus University, 8000 Aarhus C, Denmark}
\author{J.I.~Pascual}
\affiliation{CIC nanoGUNE,  20018 San Sebasti\'an-Donostia, Spain}
\affiliation{Ikerbasque, Basque Foundation for Science, 48011 Bilbao, Spain}
\author{Justin W.~Wells}
\affiliation{Dept. of Physics, Norwegian University of Science and Technology (NTNU), 4791 Trondheim, Norway.}

\begin{abstract}
Scanning tunnelling microscopy and low energy electron diffraction show a dimerization-like reconstruction in the one-dimensional atomic chains on Bi(114) at low temperatures. While one-dimensional systems are generally unstable against such a distortion, its observation is not expected for this particular surface, since there are several factors that should prevent it: One is the particular spin texture of the Fermi surface, which resembles a one-dimensional topological state, and spin protection should hence prevent the formation of the reconstruction. The second is the very short nesting vector $2 k_F$, which is inconsistent with the observed lattice distortion. A nesting-driven mechanism of the reconstruction is indeed excluded by the absence of any changes in the electronic structure near the Fermi surface, as observed by angle-resolved photoemission spectroscopy. However, distinct changes in the electronic structure at higher binding energies are found to accompany the structural phase transition. This, as well as the observed short correlation length of the pairing distortion, suggest that the transition is of the strong coupling type and driven by phonon entropy rather than electronic entropy.
\end{abstract}

\maketitle
The experimental realization of systems with reduced dimensions has often been the key to the discovery of fundamentally new physics. Of particular importance is the situation in one dimension, with its drastically enhanced significance of  electronic correlations and electron-phonon coupling \cite{Peierls:1955aa,Luttinger:1960ab}.  An attractive path to studying systems of reduced dimensionality is to create them on the surfaces of semiconducting or semimetallic substrates, as this opens the possibility to employ powerful spectroscopic techniques, such as scanning tunnelling spectroscopy (STM), angle-resolved photoemission (ARPES) and even surface-sensitive transport measurements \cite{Hofmann:2009aa,Wells:2008ab}. Many systems have been realized and studied in this way, such as metallic chains or graphene nano-ribbons on semiconductors, see e.g. Refs. \cite{Yeom:1999ab,Crain:2003aa,Tegenkamp:2008aa,Tegenkamp:2012aa,Park:2013aa,Baringhaus:2014aa,Cheon:2015aa}.

A particularly intriguing situation arises when low dimensionality is combined with an unconventional spin texture of the electronic states, as this imposes a number of restrictions on the allowed electronic instabilities \cite{Kim:2005aa,Tegenkamp:2012aa}. This combination is realized on the (114) \cite{Wells:2009aa} and (441) \cite{Bianchi:2015aa} vicinal surfaces of Bi where strongly Rashba-split surface states span the gap of a semimetallic substrate. Moreover, due to the similarity of Bi to the topological insulator Bi$_{1-x}$Sb$_x$ \cite{Teo:2008aa,Hsieh:2008aa}, all Bi surfaces have metallic surface states with a spin texture similar to that of topological insulators \cite{Hofmann:2006aa,Qi:2011aa} and several hallmark features of these states, such as the lack of backscattering and forbidden charge density waves, which were first observed on Bi surfaces \cite{Pascual:2004aa,Kim:2005aa}. 

\begin{figure}
 \includegraphics[width=\columnwidth]{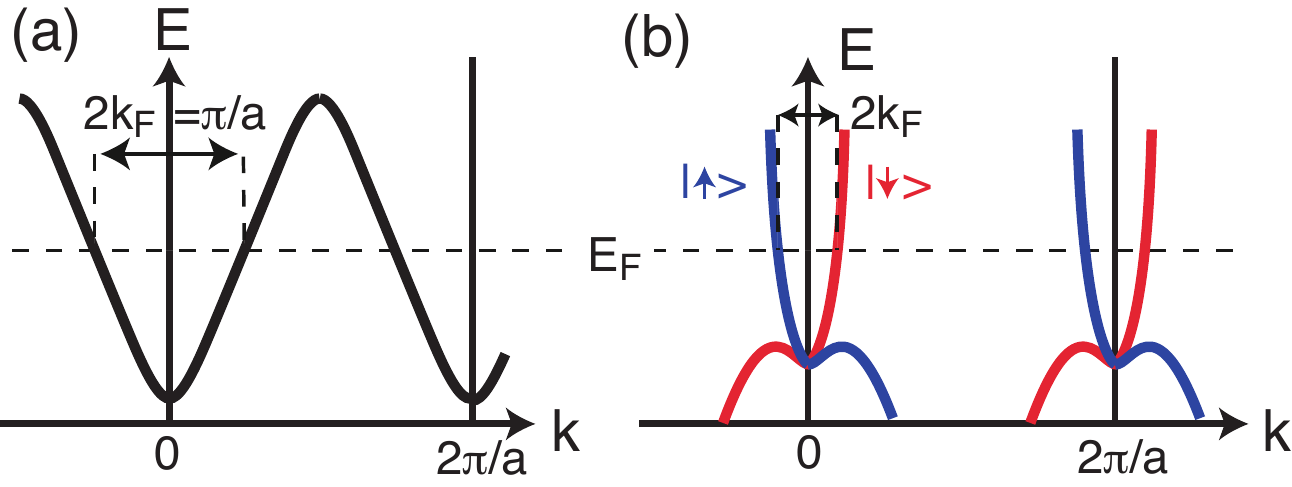}
\caption{(Color online) (a) Conventional one-dimensional electronic state at half-filling. The black dispersion is spin-degenerate. This system is sensitive to a Peierls-type instability due to the perfect nesting and the fact that the nesting vector's length corresponds to a real space periodicity of $2a$. (b) Situation on Bi(114). The spin degeneracy of the bands is lifted and while perfect nesting is still present, it takes place for a very short nesting vector and between states of opposite spin (indicated by the color of the bands and the arrows), thus protecting against a Peierls-type instability .}
\label{fig:0}
\end{figure}

The fundamental difference between a conventional one-dimensional metal at half-filling and the electronic surface state on Bi(114) is illustrated in Fig. \ref{fig:0}. A conventional electronic state in a lattice with spacing $a$ and half-filling is unstable with respect to the formation of a one-dimensional charged density wave (CDW), also called a Peierls-distortion.  Fig. \ref{fig:0}(a) illustrates this situation of ideal nesting with a nesting vector length of $2 k_F =\pi/a$, as indicated by the arrow in Fig. \ref{fig:0}(a). As this corresponds to a real space periodicity of $2a$, a Peierls-type distortion involving a periodicity doubling via pairing of the atoms in a chain leads to a new Brillouin zone boundary at the Fermi level crossings, a gap opening, and hence to an electronic energy gain in the occupied states. The situation is quite different for the surface state on Bi(114) shown in Fig. \ref{fig:0}(b). Here the state is no longer spin-degenerate and the spin texture (indicated by red and blue) and dispersion closely resemble that of a one-dimensional edge state of a two-dimensional topological insulator, the so-called quantum spin Hall effect \cite{Kane:2005ab,Murakami:2006aa,Bernevig:2006ab,Konig:2007aa};  it is for this reason that  Bi(114) has been called a `one-dimensional topological metal' \cite{Wells:2009aa}. While perfect nesting is retained, as for any one-dimensional structure,  $2 k_F \ll \pi / a$ and a pairing-type reconstruction would not be expected to open a gap at the Fermi level crossing. More importantly, the spin texture prevents the singularity in the electronic susceptibility required to drive the instability \cite{Kim:2005aa}, an effect closely related to the forbidden backscattering for such a one-dimensional state  \cite{Pascual:2004aa}.  A Peierls-distortion would thus not be expected for the case of Bi(114).

Surprisingly, as we report here, such a pairing distortion is nevertheless observed at low temperature on the quasi one-dimensional Bi(114) surface. The transition bears the signs of a strong-coupling CDW  \cite{McMillan:1977aa}, i.e. a short coherence length, large distortions, and electronic structure changes far away from the Fermi energy. These findings suggest that the transition is driven by phonon entropy, rather than electronic entropy. The observation of a strong-coupling CDW on Bi(114) illustrates that such instabilities are still possible, even though unexpected, in topological systems.

The Bi(114) surface was cleaned by sputtering with noble gas ions and annealing between 300 and 400~K.  STM measurements were performed both at a fixed temperature (5~K) and at variable  temperatures between 40~K and 300~K in two different setups. Low energy electron diffraction (LEED) and  ARPES data were collected on the SGM-3 end station of ASTRID \cite{Hoffmann:2004aa}  between 50 and 300~K. The energy resolution varied between 25~meV for the measurements at low photon energies and 65~meV for the large-scale Fermi surface maps collected with high photon energies. The angular resolution was better that 0.2$^{\circ}$.

\begin{figure}
 \includegraphics[width=\columnwidth]{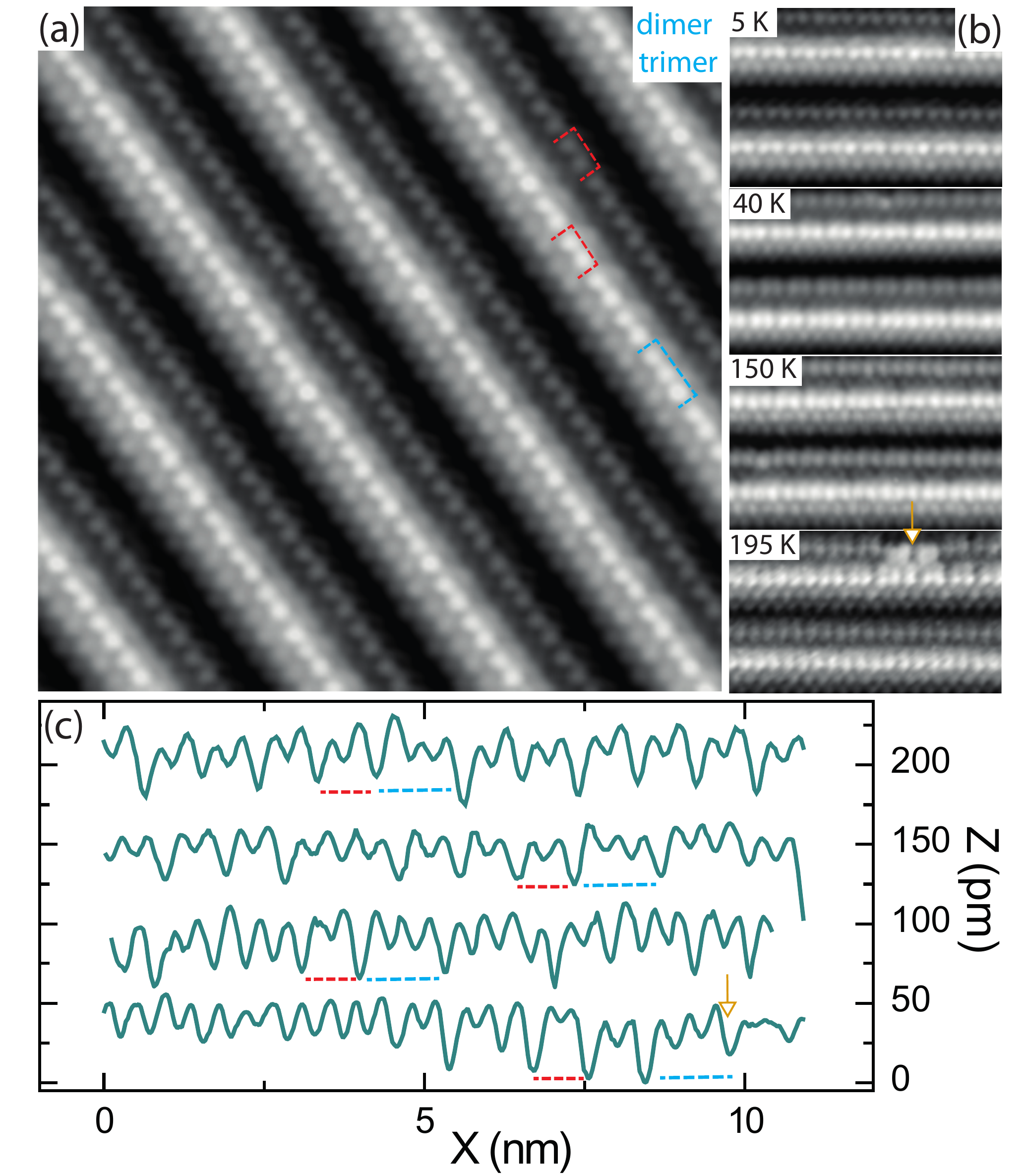}
\caption{(Color online) Dimerization distortion of the quasi one-dimensional lattice structure of Bi(114). (a) STM image at 5~K showing dimer formation in some of the  protruding atomic rows (red dashed frames), with frequently  appearing defects such as trimers (blue dashed frame). (b) STM images at higher temperatures and (c) profiles along the protruding atomic rows, measured at the indicated temperatures. At 195~K the dimerization is  lifted and only observed in the immediately vicinity of defects.  Tunnelling parameters: (a) $U=$0.2~V and and $I=$0.1 nA  (b) $U=$0.2, 1.0, -0.1, 0.25~V and $I=$0.1, 4.0, 2.0, 4.0~nA.  
All STM data were processed with the WSxM software \cite{Horcas:2007aa}.}
\label{fig:1}
\end{figure}

The atomic structure of Bi(114) consists of parallel atomic rows with an interatomic spacing of 4.54~\AA~along the rows \cite{Wells:2009aa}. A unit cell comprises several of these rows at different heights and the periodicity of the unreconstructed surface perpendicular to the rows is  14.20~\AA. For the clean surface, a reconstruction-induced doubling of this periodicity has been reported \cite{Wells:2009aa}, such that the actual periodicity perpendicular to the rows is twice this value. Fig. \ref{fig:1}(a) shows a close-up STM image of the surface at 5~K. Due to the strong corrugation, the periodicity of 28.40~\AA~perpendicular to the atomic rows is immediately visible. On close inspection, it becomes clear that the periodicity along the rows is also doubled, i.e. the atoms in the rows are not equally spaced but dimerized. A few of these dimers are emphasized in the figure by red frames. This reconstruction does not only affect the atoms in the top row but also those in deeper lying rows. The average interatomic distance in the dimers is 4.15~$\pm$~0.1\AA, thus corresponding to $\sim$~20~pm atomic displacement from the equilibrium position. This is a substantial fraction of the equilibrium spacing, larger than typically found in Peierls systems \cite{Pouget:2016aa}.  While dimerization is dominant on a short length scale, long range ordering is poor. Indeed,  defects  such as trimers are found roughly every few dimers in the row (also indicated by a blue frame in the figure). Moreover, the dimerization appears to be a strongly local phenomenon within each row, as no correlations between dimer positions in neighbouring atomic rows are evident.

We explored the temperature dependence of the dimerization along the atomic rows.  Fig. \ref{fig:1}(b) and \ref{fig:1}(c) compare  STM images and height profiles  measured at increasing temperatures. While at 40 K the dimerization is still fully intact, at 150~K it starts vanishing, and  is nearly absent at 195~K, with the exception of the immediate vicinity of structural defects (indicated by an arrow) that appear to serve as a seed for the dimerization.   Thus, the dimerization is a low temperature phase, with transition temperature somewhere between 150 and 195~K. 

\begin{figure}
 \includegraphics[width=\columnwidth]{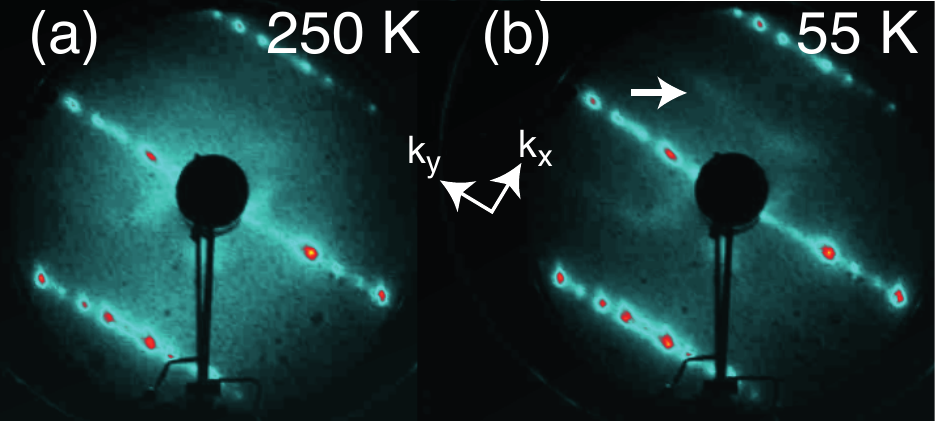}
\caption{(a) and (b) LEED patterns collected above and below the temperature of the  phase transition, respectively. The arrow in (b) shows the additional streaks induced by the periodicity doubling along the atomic chains. The electron kinetic energy is 27.2~eV. }
\label{fig:2}
\end{figure}

The reconstruction should also be observable in diffraction experiments, even though the coherence length is very short, as seen from the STM data.  Fig. \ref{fig:2} shows  LEED data taken at 250 and 55~K, i.e. well above and below the transition temperature range determined by STM.  The LEED patterns consist of well-separated rows of closely placed sharp spots. The distance between the spots along the rows corresponds to the reciprocal lattice distance perpendicular to the chains ($2 \pi / 28.4 $~\AA$^{-1}$), while the distance between the rows corresponds to the interatomic distance in the (unreconstructed) chains ($2 \pi / 4.54 $~\AA$^{-1}$). In the low-temperature image, weak streaks of intensity are observed half-way between the rows  (indicated by an arrow) and thus consistent with a dimerized structure. The features are  poorly defined in the direction perpendicular to the atomic rows, indicating a  low degree of coherence between dimers in neighbouring unit cells, also consistent with the STM results. Temperature-dependent LEED results could potentially yield more accurate information about the character of the phase transition and the transition temperature. Unfortunately, such information is difficult to extract due to the weakness of the superstructure feature and the strong, and strongly temperature-dependent, background which is caused by the low Debye temperature of Bi \cite{Monig:2005aa}. 

\begin{figure}
 \includegraphics[width=\columnwidth]{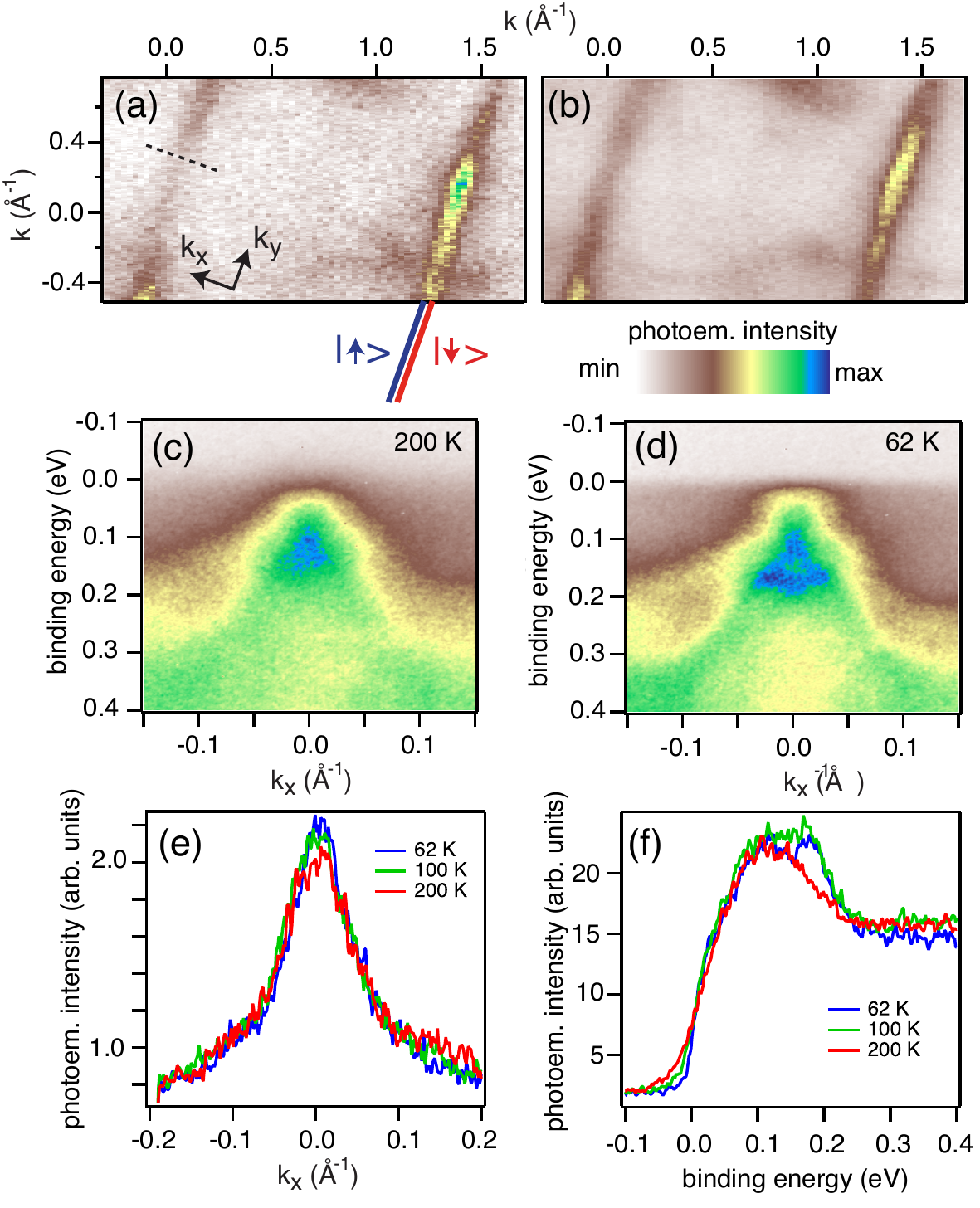}
\caption{(Color online) (a),(b) Photoemission intensity in the low-temperature phase at $T=$60~K in a 20~meV window around the Fermi energy and a binding energy of 170~meV, respectively. The sketched continuation of the Fermi surface illustrates that the observed line-like intensity is comprised of two unresolved spin-polarized Fermi level crossings \cite{Wells:2009aa}, see also Fig. \ref{fig:0}(b). The photon energy is $h\nu = 70$~eV. The dashed line shows the direction of the data in (c)-(e). (c), (d) Photoemission intensity above and below the Peierls transition (at $T=$200 and 62~K), respectively, along the dashed black line in (a). (e) Momentum distribution curves at the Fermi energy along the dashed line in (a) as a function of temperature. (f) Energy distribution curves along the crossing point of the dashed line in (a) and the Fermi surface at $k_x=0$, i.e. through the center of the images in (c) and (d). $h\nu = 17$~eV in (c)-(f). }
\label{fig:3}
\end{figure}

A clearer picture of the temperature-dependent phase transition and its influence on the surface electronic structure is given by ARPES. Fig. \ref{fig:3}(a) shows the photoemission intensity at the Fermi level at 60~K. It is dominated by intense lines in the direction perpendicular to the atomic rows. As has been shown by spin-resolved photoemission, these intense lines are actually caused by the two unresolved spin-polarized Fermi surface elements due to the spin-split surface state in Fig. \ref{fig:0}(b). This unresolved Fermi contour is indicated as a sketch, extending the observed Fermi contour in the figure. Apart from the intense linear features, the photoemission intensity shows some weak structures that can be assigned to bulk states \cite{Bianchi:2015aa}.

Since the Fermi surface of Fig. \ref{fig:3}(a) is measured in the dimerized phase, one might expect replicas of the Fermi surface lines in between the two intense lines, similar to the additional weak streaks in between the lines of densely spaced spots in the LEED image of Fig. \ref{fig:2}(b). Such replicas are not observed, neither at the Fermi surface in Fig. \ref{fig:3}(a) nor for the higher binding energy of 170~meV in Fig. \ref{fig:3}(b) that will discussed further below.

Fig. \ref{fig:3}(c) and (d) show the electronic structure of the one-dimensional states above and below the  dimerization transition temperature, respectively, as a function of binding energy in the $k_x$-direction perpendicular to the one-dimensional Fermi surface, along a cut indicated by the black dashed line in Fig. \ref{fig:3}(a). No spectroscopic signature of the transition is observed near the Fermi energy. Indeed, when taking momentum distribution curves (MDCs) through a temperature-dependent series of similar data sets, 
no significant changes can be observed (see Fig. \ref{fig:3}(e)), except for a  minor sharpening of the MDC peak at low temperatures, as expected due to electron-phonon coupling \cite{Gayone:2003aa,Hofmann:2009ab}. 

However, there is a significant change in the electronic structure rather far away from the Fermi energy, at around 150~meV, where a single intense feature at 110~meV in the high temperature phase of Fig. \ref{fig:3}(c) splits into two peaks, one at essentially staying the same binding energy and one moved to a higher energy of 175~meV. This spectral change is especially well seen in energy distribution curves through the center of a series of temperature-dependent data \cite{SMAT} shown in Fig. \ref{fig:3}(f). These changes in the electronic structure at high binding energy do not lead to observable replicas in the constant energy surfaces either, as seen in Fig. \ref{fig:3}(b) which shows the photoemission intensity at a binding energy of 170~meV.

It is surprising that the strong lattice distortion observed by STM  is only accompanied by spectral changes at a high binding energy.  This observation is not consistent with a Fermi surface nesting-driven weak coupling CDW,  such as the Peierls distortion in Fig. \ref{fig:0}(b), as this would require a gap opening at the Fermi energy. On the other hand,  an electronic structure change in which states are shifted to a substantially higher binding energy can certainly  explain the stabilization of the reconstructed phase. Indeed, the splitting observed in Fig. \ref{fig:3}(f) resembles a bonding/antibonding splitting where the antibonding state remains at a similar binding energy but the formation of the bonding state is accompanied by an energy gain. This gives a qualitative explanation for the stability of the pairing distortion at low temperatures, but it does not explain why the pairing should be lifted at higher temperature.   After all, changes in the electron occupation happen only in a narrow window around the Fermi energy compared to the binding energy in question here, and there is no gap around the Fermi energy that would be bridged by temperature-induced electronic excitations. 

The observed changes in the electronic structure thus clearly do not support the scenario of a conventional Peierls transition that is driven by electronic entropy, but several experimental facts are consistent with a strong coupling scenario that is driven by phonon entropy \cite{McMillan:1977aa,Gruner:1994aa}. In such a transition the energy gain is still electronic but the size of the gap is much larger than typical phonon energies. In fact, the gap remains large across the transition. A strong coupling-CDW can further be expected to show a short coherence length and considerable atomic displacements, also consistent with the observations here. 

An argument against the strong-coupling CDW scenario is that the local order would be expected to be present well above the transition temperature and only the long-range order would be completely lost. The STM results of Fig. \ref{fig:1}(b) could appear to suggest that no short range order is present at 195~K  and above. We must keep in mind, however, that strong fluctuations are to be expected and that STM, being a slow technique, only measures the \emph{average} position of the atomic motion and might thus not be able to show the preserved local dimerization at high temperatures. This actually explains why close to defects, dimerization can still be observed frozen in the  195~K images. Indeed, LEED patterns taken at 200 and even 250~K do still show very weak signatures of the superstructure (see Fig. \ref{fig:2}(a) and Ref. \cite{SMAT}). A strong-coupling CDW would not be uncommon for a one-dimensional surface structure. A similar mechanism has also been suggested for the CDW of In nanowires on Si(111) \cite{Wippermann:2010aa} which, however, also shows a pronounced gap opening \cite{Yeom:1999ab,Frigge:2017aa}. Most importantly, the strong coupling picture would explain why a temperature-dependent distortion can appear in a system with this particular spin texture, for which it would otherwise not be expected.

Bi(114) has recently been found to show a pronounced excitation of   coherent phonons that can be launched by an ultrashort light pulse at room temperature. The observed frequency of 0.72~THz was assigned to high phonon density of states along the $K-X$ direction at the bulk Brillouin zone boundary, i.e. along the direction of the atomic rows of Bi(114) \cite{Leuenberger:2013aa}. Our unexpected finding of a CDW transition below room temperature could explain the origin of the surface's strong coherent response at this frequency as a transient excitation of the incipient CDW, an interpretation that was previously excluded because of the spin texture of the surface Fermi contour \cite{Leuenberger:2013aa}.

In conclusion, we have reported the observation of a  dimerization transition below 195~K on Bi(114). Such a transition is highly unexpected, even in the presence of a perfectly nested one-dimensional Fermi contour, because it appears to be forbidden by the spin texture of the states and the non-matching nesting vector $2 k_F$. Indeed, the transition does not involve the states near the Fermi energy but spectral changes at higher binding energy are observed. This, as well as the short coherence length, support the interpretation of the low-temperature state as a strong-coupling CDW, illustrating that such transitions are still possible in topological systems.

We gratefully acknowledge stimulating discussions with Kai Rossnagel, Matthias Hengsberger and  J\"urg Osterwalder. This work was supported by the Danish Council for Independent Research, Natural Sciences under the Sapere Aude program (Grant No. DFF-4002-00029) and by VILLUM FONDEN via the center of Excellence for Dirac Materials (Grant No. 11744). JW acknowledges support from the Institute for Storage Ring Facilities, Aarhus University, during the beamtime for this project, and from the Research Council of Norway through project number 250985/F20 ``FunTopoMat'' in the \textit{fripro} program. A.J.M.-G. acknowledges funding from the Spanish MINECO through the Juan de la Cierva program. J.M.G.-R acknowledges financial support from the Spanish MINECO under project number MAT2016-77852-C2-2-R.
%
%

\end{document}